\def\figsize{9.5cm}

\def\rn{}
\def\nn#1 #2{#2. #1}				
\def\nnn#1 #2 #3{#2. #3. #1}			
\def\nnnn#1 #2 #3 #4{#2. #3. #4 #1}		
\def\nnnnn#1 #2 #3 #4 #5{#2. #3. #4 #5. #1}	
\def\dualand{ and\hbox{ }}				
\def\multiand{, and\hbox{ }}				
\def\rf#1;#2;#3;#4;#5 {{\frenchspacing\par\rn#1, #3 {\bf #4}, #5 (#2). \par}}
\def\rg#1;#2;#3;#4;#5;#6 {{\frenchspacing\par\rn#1, #3 {\bf #4}, #5 (#2). \par}}
\def\rfbook#1;#2;#3;#4;#5 {{\frenchspacing\par\rn#1, {\it #3} (#5, #4, #2).\par}}
\def\rfprep#1;#2;#3 {{\par\frenchspacing\rn#1, #3 (#2).\par}}
\def\rfproc#1;#2;#3;#4;#5;#6 {{\frenchspacing\par\rn#1 #2, in {\it #3}, ed. #4 (#5: #6)\par}}
\def\rfprocp#1;#2;#3;#4;#5;#6;#7 {{\frenchspacing\par\rn#1 #2, in {\it #3}, ed. #4 (#5: #6), p#7\par}}

\def\rg#1;#2;#3;#4;#5;#6 {\par\rn#1 #2, {\it #3}, {\bf #4}, #5 (``#6'') \par}
\def\rf#1;#2;#3;#4;#5 {\par\rn#1, {\it #3}, {\bf #4}, #5 (#2)\par}
\def\rfbook#1;#2;#3;#4;#5 {{\frenchspacing\par\rn#1, {\it #3} (#4: #5, #2)\par}}
\def\rfproc#1;#2;#3;#4;#5;#6 {{\frenchspacing\par\rn#1 #2, in {\it #3}, ed. #4 (#5: #6)\par}}
\def\rfprocp#1;#2;#3;#4;#5;#6;#7 {{\frenchspacing\par\rn#1 #2, in {\it #3}, ed. #4 (#5: #6), p#7\par}}
\def\rfprep#1;#2;#3  {{\par\rn#1, #3, #2\par}}
\def\rfprepp#1;#2;#3 {{\par\rn#1 #2, #3\par}}

\def\eV{{\rm eV}}

\def\etal{{\frenchspacing\it et al.}}
\def\ie{{\frenchspacing\it i.e.}}

\def\beq#1{\begin{equation}\label{#1}}
\def\eeq{\end{equation}}
\def\beqa#1{\begin{eqnarray}\label{#1}}
\def\eeqa{\end{eqnarray}}
\def\eq#1{equation~(\ref{#1})}
\def\Eq#1{Equation~(\ref{#1})}

\def\fig#1{Figure~\ref{#1}}
\def\Fig#1{Figure~\ref{#1}}

\def\spose#1{\hbox to 0pt{#1\hss}}
\def\simlt{\mathrel{\spose{\lower 3pt\hbox{$\mathchar"218$}}
     \raise 2.0pt\hbox{$\mathchar"13C$}}}
\def\simgt{\mathrel{\spose{\lower 3pt\hbox{$\mathchar"218$}}
     \raise 2.0pt\hbox{$\mathchar"13E$}}}
\def\simpropto{\mathrel{\spose{\lower 3pt\hbox{$\mathchar"218$}}
     \raise 2.0pt\hbox{$\propto$}}}

\def\ed{\end{document}}

\def\Om{\Omega_m}

\def\od{\omega_d}

\def\om{\omega_{\rm m}}
\def\on{\omega_\nu}

\def\fn{f_\nu}

\def\Mnu{M_\nu}

\def\beq#1{\begin{equation}\label{#1}}
\def\eeq{\end{equation}}
\def\beqa#1{\begin{eqnarray}\label{#1}}
\def\eeqa{\end{eqnarray}}
\def\eq#1{equation~(\ref{#1})}
\def\Eq#1{Equation~(\ref{#1})}

\def\ignore#1{}

\def\simless{\mathbin{\lower 3pt\hbox
        {$\,\rlap{\raise 5pt\hbox{$\char'074$}}\mathchar"7218\,$}}} 
\def\simgreat{\mathbin{\lower 3pt\hbox
        {$\,\rlap{\raise 5pt\hbox{$\char'076$}}\mathchar"7218\,$}}} 

\documentclass[twocolumn,amsmath,nofootinbib]{revtex4} 
\usepackage{url}
\begin{document}
\input{epsf.sty}




\def\mit{1}
\def\penn{2}

\def\affilmrk#1{$^{#1}$}
\def\affilmk#1#2{$^{#1}$#2;}


\title{Cosmological neutrino bounds for non-cosmologists}

\author{
Max Tegmark\affilmrk{\mit,\penn}
}
\address{
\affilmk{\mit}{Dept. of Physics, Massachusetts Institute of Technology, 
Cambridge, MA 02139}\\
\affilmk{\penn}{Department of Physics, University of Pennsylvania,
Philadelphia, PA 19104, USA}
}


\date{February 18, 2004. To appear in {\it Neutrino Physics}, proceedings of the Nobel Symposium 2004\\ 
zzEnk{\"o}ping, Sweden, August 19-24, 2004,
Eds. L. Bergstr{\"o}m, O. Botner, P. Carlson, P.O. Hulth \& T. Ohlsson}

\begin{abstract}
I briefly review cosmological bounds on neutrino masses and the underlying gravitational physics  
at a level appropriate for readers outside the field of cosmology.
For the case of three massive
neutrinos with standard model freezeout, the current 95\% upper limit on the sum of their masses is 0.42 eV.
I summarize the basic physical mechanism making matter clustering such a sensitive probe of massive neutrinos.
I discuss the prospects of doing still better in coming years using tools such as lensing tomography, approaching a
sensitivity around 0.03 eV. Since the lower bound from atmospheric neutrino oscillations is around 
0.05 eV, upcoming cosmological measurements should detect neutrino mass 
if the technical and fiscal challenges can be met.
\end{abstract}

\keywords{large-scale structure of universe 
--- galaxies: statistics 
--- methods: data analysis}

\pacs{98.80.Es}
  
\maketitle



\setcounter{footnote}{0}

\section{Introduction}

In the last few years, an avalanche of new cosmological data has revolutionized our ability
to measure key cosmological parameters. Measurements of the cosmic microwave background (CMB),
galaxy clustering, gravitational lensing, the Lyman alpha forest, cluster abundances and type Ia supernovae
paint a consistent picture where the cosmic matter budget is about 5\% ordinary matter, 25\% dark matter,
70\% dark energy, less than 1\% curvature and less than 5\% massive neutrinos
\cite{Spergel03,sdsspars,sdsslyaf,2dfpars,sdssbump}. 
The cosmic initial conditions are consistent with approximately scale-invariant inflation-produced adiabatic fluctuations,
with no evidence yet for primordial gravitational  waves \cite{Spergel03,sdsspars,sdsslyaf,2dfpars,sdssbump}.

How precisely do such cosmological observations give us information about neutrino masses, and how should the current limits be interpreted?
For detailed discussion of post-WMAP astrophysical neutrino constraints, see 
\cite{Spergel03,sdsspars,sdsslyaf,Hannestad0303076,ElgaroyLahav0303089,BashinskySeljak03,Hannestad0310133,Hannestad0411475,Hannestad0412181}
and in particular the excellent and up-to-date reviews \cite{Hannestad0404239,ElgaroyLahav0412075}.
For a review of the theoretical and experimental situation, see \cite{King03}.
The purpose of this symposium contribution is merely to provide a brief summary of the constraints and the underlying physics at 
a level appropriate for readers outside of the field of cosmology.

\section{The physics underlying cosmological neutrino bounds}

\begin{figure} 
\centerline{\epsfxsize=\figsize\epsffile{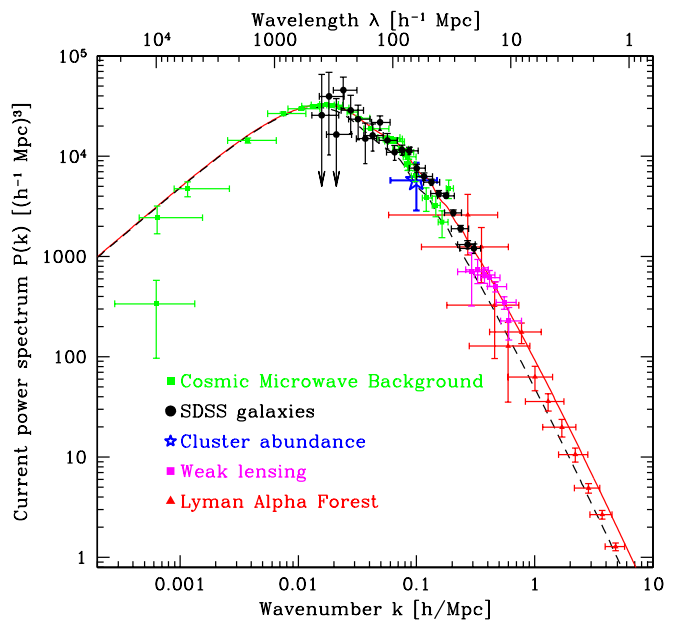}}
\caption[1]{\label{kspace_fig}\footnotesize%
Cosmological constraints on the current matter power spectrum $P(k)$
reprinted from \cite{sdsspower}. See \cite{sdsspower} for details about the 
modeling assumptions underlying this figure.
The solid curve shows the theoretical prediction for  
a ``vanilla'' flat scalar scale-invariant model with matter density
$\Om=0.28$, Hubble parameter $h=0.72$ and baryon fraction $\Omega_b/\Omega_m=0.16$.
The dashed curve shows that replacing 7\% 
of the cold dark matter by neutrinos, corresponding 
to a neutrino mass sum $\Mnu=1$ eV, suppresses small-scale power by about a factor of two.
}
\end{figure}

Why do cosmological observations place strong bounds on neutrino masses?
The short answer is that neutrinos affect the growth of cosmic clustering and 
this clustering can be accurately measured (\fig{kspace_fig}).

The CMB tells us that the Universe used to be almost perfectly uniform spatially, with density variations from place to place only
at the level of $10^{-5}$. Gravitational instability caused these tiny density fluctuations to grow in amplitude into the 
galaxies and the large-scale structure that we observe around us today. The reason for this growth is simply that gravity is an attractive force:
if the density at some point exceeds the mean density by some relative amount $\delta$, then mass will be pulled in from surrounding regions
and $\delta$ increases over time.
A classic result is that if all the matter contributing to the cosmic density is able to cluster 
(like dark matter or ordinary matter with negligible pressure), then fluctuations grow as the 
cosmic scale factor $a$  \cite{KolbTurnerBook}:
\beq{GrowthEq}
\delta\propto a,
\eeq
\ie, fluctuations double in amplitude every time the Universe doubles it linear size $a$.

If some fraction of the matter density is gravitationally inert and unable to cluster, the fluctuation growth will clearly be slower.
If only a fraction $\Omega_*$ can cluster, then \eq{GrowthEq} is generalized to \cite{BondEfstathiouSilk80}: 
\beq{GrowthEq2}
\delta\propto a^p,
\eeq
where
\beq{pEq}
p = {\sqrt{1+24\Omega_*}-1\over 4}\approx\Omega_*^{3/5}
\eeq
and the approximation in the last step is surprisingly accurate.
Such gravitationally inert components can include dark energy and (on sufficiently small scales) photons and 
neutrinos.
Early on, the cosmic density was completely dominated by photons, so $p\approx 0$ and fluctuations essentially did not start growing
until the epoch of matter-domination (MD). 
At recent times, the cosmic density has become dominated by dark energy $\Lambda$,
causing fluctuations to gradually stop growing after
a net growth factor of about $a_{\Lambda{\rm D}}/a_{\rm MD}\approx 4700$ \cite{anthroneutrino}.

Massive nonrelativistic neutrinos are unable to cluster on small scales because of their high velocities.
Between matter domination and dark energy domination, they constitute a roughly constant 
fraction $\fn=1-\Omega_*$ of the matter density. \Eq{GrowthEq2} therefore gives a
net fluctuation growth factor 
\beq{NetGrowthEq}
\left({a_{\Lambda{\rm D}}\over a_{\rm MD}}\right)^p\approx 4700^p\approx 4700^{(1-\fn)^{3/5}}\approx 4700 e^{-4\fn}
\eeq
from matter-domination (MD) until today, where we have assumed $\fn\ll 1$ in the last step.
We see that the basic reason that a small neutrino fraction has a large effect is simply that $4700$ is a large
number, so that a small change in the exponent $p$ makes a noticeable difference. 

A key point to remember is that what mattered above was the neutrino ${\it density}$, specifically
the fractional contribution $\fn$ of neutrinos to the total density. To translate
observational constraints on the neutrino mass density into constraints on neutrino masses,
we need to know the neutrino number density. 
Assuming that this number density is determined by standard model neutrino freezeout 
gives a number density around 112/cm$^3$ and \cite{KolbTurnerBook}\footnote{The neutrino energy density must be
very close to the standard freezeout density \cite{synch1,synch2,synch3}, given
the large mixing angle solution to the solar neutrino problem
and near maximal mixing from atmospheric results--- see \cite{Kearns02,Bahcall03} for up-to-date reviews.  
Any substantial asymmetries in neutrino density from the standard value would
be ``equilibrated'' and produce a primordial $^4$He abundance inconsistent with
that observed.
}
\beq{fnEq}
\fn\approx{\Mnu\over\om\times 94.4 \eV} \approx{\Mnu\over 14 \eV},
\eeq
where
\beq{MnuDefEq}
\Mnu\equiv \sum_{i=1}^3 m_\nu^i
\eeq
is the sum of the three neutrino masses
and
$\om=h^2\Om\approx 0.15$ is the measured matter density in units of $1.8788\times 10^{-26}$kg$/$m$^3$ \cite{sdsspars}.

The power spectrum $P(k)$ shown in \fig{kspace_fig} is the variance of the fluctuations $\delta$ in Fourier space,
so massive neutrinos suppress it by the same factor as it suppresses $\delta^2$, \ie, by a factor \cite{neutrinos}: 
\beq{PowerSuppressionEq}
{P(k;\fn)\over P(k;0)} \approx e^{-8\fn}.
\eeq
This means that a neutrino mass sum $\Mnu=1 eV$ cuts the power roughly in half on the small scales where neutrinos 
cannot cluster.

The length scale below which neutrino clustering is strongly suppressed is called the neutrino free-streaming scale,
and roughly corresponds to the distance neutrinos have time to travel while the Universe expands by a factor of two.
Intuitively, neutrinos clearly will not cluster in an overdense clump so small that its escape velocity is much 
smaller than the typical neutrino velocity.
On scales much larger than the free-streaming scale, on the other hand, neutrinos cluster just as cold dark matter
and give $\Omega_*=1$ and $p=1$ above. This explains the effect of neutrinos on the power
spectrum seen in \fig{kspace_fig}: suppression only on small scales, thereby changing the overall shape of $P(k)$
in a characteristic way.
This shape change distinguishes the neutrino effect from that of dark energy, which lowers $\Omega_*$ and hence
the power spectrum amplitude on {\it all} scales by the same factor, preserving the shape of $P(k)$.


\section{What are the constraints?}

\begin{figure} 
\centerline{\epsfxsize=\figsize\epsffile{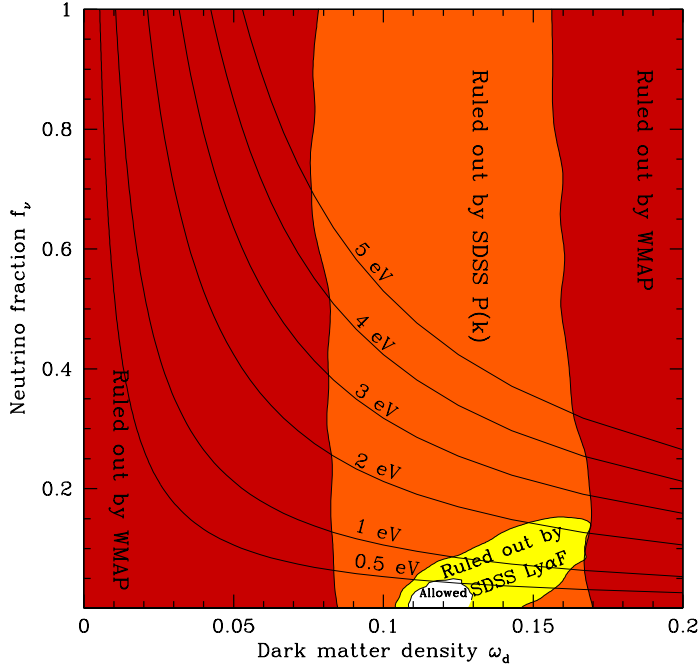}}
\caption[1]{\label{2d_odfn_7parnu_fig}\footnotesize%
95\% constraints in the $(\od,\fn)$ plane, reprinted from \cite{sdsspars} and \cite{sdsslyaf}. 
The shaded red/dark grey region is ruled out by WMAP CMB observations alone.
The shaded orange/grey region is ruled out when adding SDSS galaxy clustering information \cite{sdsspars}
and the yellow/light grey region is ruled out when including SDSS Lyman $\alpha$ Forest information
as well \cite{sdsslyaf}.
The five curves correspond to $M_\nu$, the sum of the neutrino masses,
equaling 1, 2, 3, 4 and 5 eV, respectively --- barring sterile neutrinos,
no neutrino can have a mass exceeding $\sim M_\nu/3.$ 
}
\end{figure}
Cosmological observations have thus far produced no convincing detection of neutrino mass, but strong upper limits
as illustrated in \Fig{2d_odfn_7parnu_fig}.
This figure shows that 
the WMAP CMB-measurements alone \cite{Bennett03} tell us almost nothing about neutrino masses and are consistent 
with neutrinos making up 100\% 
of the dark matter. Rather, the power of WMAP is that it constrains other cosmological parameters so strongly 
that it enables large-scale structure data to measure the small-scale $P(k)$-suppression 
that massive neutrinos cause.
Combining WMAP with Sloan Digital Sky Survey (SDSS) \cite{York00} galaxy clustering measurements \cite{sdsspower} 
gives the most favored value $\Mnu=0$ and the
95\%  upper limit $\Mnu<1.7\>$eV \cite{sdsspars}.
Including information about SDSS or 2dFGRS \cite{Percival01} galaxy bias tightens this bound to $\Mnu<0.6-0.7$ eV \cite{Spergel03,sdssbias}.
Including SDSS measurements of the so-called Lyman $\alpha$ forest (intergalactic gas backlit by quasars)
further tightens the bound to $\Mnu<0.42$ eV (95\%).

These upper limits are complemented by the lower limit from neutrino oscillation experiments. 
Atmospheric neutrino oscillations
show that there is at least one
neutrino (presumably mostly a linear combination of 
$\nu_\mu$ and $\nu_\tau$) whose mass exceeds a lower limit
around $0.05\>$eV \cite{Kearns02,King03}.
Thus the atmospheric neutrino data corresponds to a lower limit 
$\on\simgt 0.0005$, or
$\fn\simgt 0.004$.
The solar neutrino oscillations occur at a still smaller mass scale, perhaps around $0.008\>$eV
\cite{sno0309004,King03,Bahcall03}.
These mass-splittings 
are substantially smaller than 0.42 eV, suggesting that all three mass eigenstates would
need to be almost degenerate for neutrinos to weigh in near our upper limit.
Since sterile neutrinos are disfavored from being thermalized in the early
universe \cite{fournu1,fournu2}, it can be assumed that only three neutrino flavors are
present in the neutrino background; this means that none of the three neutrinos
can weigh more than about $0.42/3 = 0.14$ eV.  The mass of the heaviest neutrino
is thus in the range $0.05-0.14$ eV.

 
A caveat about non-standard neutrinos is in order. 
As mentioned above, the cosmological constraints to first order probe
only the {\it mass density} of neutrinos, $\rho_\nu$,
which determines the small-scale power suppression factor,
and the {\it velocity dispersion}, which determines the scale below which 
the suppression occurs. For the low mass range we have discussed,
the neutrino velocities are high and the suppression occurs on 
all scales where SDSS is highly sensitive. We thus measure only the neutrino 
mass density, and our conversion
of this into a limit on the mass sum assumes that the
neutrino number density is known and given by the standard model
freezeout calculation. In more general scenarios with sterile
or otherwise non-standard neutrinos where the freezeout abundance 
is different, the conclusion to take away is 
an upper limit on the total light neutrino mass density
of $\rho_\nu < 4.8\times 10^{-28}$kg/m$^3$ (95\%).
To test arbitrary nonstandard models, a future challenge will be to
independently measure both the mass density and the velocity dispersion,
and check whether they are both consistent with the same value of $\Mnu$.

\section{Outlook}

Although cosmological neutrino bounds have recently improved dramatically,
there is ample room for further improvement in the near and intermediate future.
The basic reason for this is that the weakest link in current constraints is their dependence
on other cosmological parameters. For instance, the galaxy clustering constraints cannot directly exploit 
the dramatic effect of neutrinos on the {\it amplitude} of the small-scale power spectrum shown in \fig{kspace_fig},
merely the slight change in its {\it shape}, and this shape change can be partially mimicked by
changing other cosmological parameters such as the spectral index from inflation which effectively tilts the $P(k)$.
The reason for this shortcoming is that galaxy surveys do not measure the clustering amplitude of matter directly, 
merely the clustering amplitude of luminous matter (galaxies), which is known to differ by a constant factor 
that is measured empirically.

Weak gravitational lensing bypasses this shortcoming. Light from distant galaxies or CMB patterns is deflected in a measurable way
by the gravitational pull of {\it all} intervening matter, regardless whether it is luminous or dark, 
baryonic or non-baryonic, allowing the matter power spectrum $P(k)$ to be measured in a clean assumption-free way.
This booming field has the potential to attain a neutrino mass sensitivity of 0.03 eV or better \cite{weaklensnu1,weaklensnu2,weaklensnu3}.
Since the lower bound from atmospheric neutrino oscillations is around 
0.05 eV, upcoming cosmological measurements should detect neutrino mass 
if the technical and fiscal challenges can be met.

\bigskip
{\bf Acknowledgements:}
The author wishes to thank Ang{\'e}lica de Oliveira-Costa for helpful comments and 
Uro\v s Seljak and Matias Zaldarriaga for publishing the CMBFast software \cite{cmbfast},
which was used for the above figures.

This work was supported by NASA grant NAG5-11099,
NSF CAREER grant AST-0134999, and fellowships from the David and Lucile
Packard Foundation and the Research Corporation.




\end{document}